\documentclass[%
 aip,
 amsmath,amssymb,
 reprint,%
numbers,
]{revtex4-1}

\usepackage{graphicx}

\usepackage{dcolumn}
\usepackage{bm}

\usepackage[utf8]{inputenc}
\usepackage[T1]{fontenc}
\usepackage{mathptmx}
\usepackage{etoolbox}
\usepackage{xcolor}
\usepackage{braket}
\usepackage{svg}
\usepackage{siunitx}
\makeatletter
\def\@email#1#2{%
 \endgroup
 \patchcmd{\titleblock@produce}
  {\frontmatter@RRAPformat}
  {\frontmatter@RRAPformat{\produce@RRAP{*#1\href{mailto:#2}{#2}}}\frontmatter@RRAPformat}
  {}{}
}%
\makeatother

\begin{document}

\preprint{APL/123-QED}

\title[]{Distribution of entangled photons from an AlGaAs Bragg-reflection waveguide over an actively stabilized 28 km urban telecom fiber link}

\author{Akriti Raj}
\author{Tobias Bauer}%
\author{Christian Haen}%
\author{David Lindler}%
\affiliation{Fachrichtung Physik, Universität des Saarlandes, Campus E2.6, Saarbrücken, 66123, Germany}
\affiliation{Zentrum für Quantentechnologien (QuTe), Universität des Saarlandes, Campus,
Saarbrücken, 66123, Germany}
\author{Quankui Yang}%
\author{Thorsten Passow}%
\author{Marko Haertelt}%
\affiliation{ 
Fraunhofer-Institut für Angewandte Festkörperphysik IAF, Tullastr. 72, Freiburg, 79108, Germany}
\author{Jürgen Eschner}%
\author{Christoph Becher}
\email{christoph.becher@uni-saarland.de}%
\affiliation{Fachrichtung Physik, Universität des Saarlandes, Campus E2.6, Saarbrücken, 66123, Germany}
\affiliation{Zentrum für Quantentechnologien (QuTe), Universität des Saarlandes, Campus,
Saarbrücken, 66123, Germany}
\date{\today}

\begin{abstract}
The generation and distribution of single photons and entangled photon pairs is central to the definition of photonic quantum bits for applications in recent quantum technologies. Semiconductor sources are particularly attractive in this context, as they are compact, operate at room temperature, and are realized in a material platform that allows monolithic integration of the pump laser on the same chip. In our work, we employ an AlGaAs Bragg reflection waveguide as the non-linear medium to create such a quantum source, producing polarization entangled qubits at telecommunication wavelengths, coupled directly into a single-mode fiber, with a high pair generation rate $ > 10^7$/s/mW. We demonstrate high heralded single photon purity with a photon correlation function $g^{(2)}(0) < 10^{-4}$. To confirm the entanglement of the generated photon pairs, we determine the fidelity ($>95\,\%$) and purity ($>91\,\%$) with respect to the $|\Psi^{+}\rangle$ Bell state. For demonstration of a field test, we transmit the photon pairs through a 28$\,$km urban deployed telecom fiber link, where active stabilization compensates the environmentally induced polarization drift. The high pair generation rate and robustness of the source allow the quality of entanglement to be largely preserved despite the loss and  residual instability of the installed fiber, providing a realistic assessment of how such a source would perform in an operational quantum network. These results show that a fiber-coupled semiconductor waveguide source can deliver entanglement of sufficient quality for quantum communication outside a laboratory environment, and constitute a step towards compact, deployable entanglement sources for real-world quantum networks.
\end{abstract}

\maketitle

\begin{quotation}
Quantum light sources are essential resources for quantum technological applications. In this paper, we realize such a source using AlGaAs Bragg reflection waveguides. Due to its high non-linear coefficient, ability to operate at room temperature, and non-birefringent characteristics, AlGaAs proves to be an excellent material choice. Here, we test the produced photons to validate their characteristics as single photons and as polarization-entangled photon pairs. We assess the relevant metrics by measuring the second order correlation function as well as fidelity and purity with respect to the $|\Psi^{+}\rangle$ Bell state. The polarization-entangled photons are distributed over a 28$\,$km dark fiber link. The link consists of a commercial fiber, deployed both underground and aerial, includes several patching points, and exhibits a transmission loss of  $\approx$ 19$\,$dB. After transmission over the actively stabilized fiber link, the photons retain their entanglement. This demonstrates the generation and robust distribution of entangled photonic qubits from a source that is particularly suitable for practical applications.    
\end{quotation}

\section{\label{sec:level1}Introduction}
Many implementations of quantum technologies rely on non-classical light sources that produce single photons or entangled photon pairs. For an easy integration of these sources for long distance applications properties such as high generation rate and wavelength within the low absorption or dispersion region (C band or O band) of standard telecom fiber are important \cite{couteau2023applications, wang2025scalable}. At present, the techniques used to generate such photons can be broadly classified into deterministic and probabilistic  \cite{aharonovich2016solid}. Deterministic sources rely on exciting a quantum emitter with an external source, which then releases a single photon upon relaxation. A wide range of such emitters is being actively investigated, including single atoms and trapped ions \cite{tissot2026hybrid, liu2026long} , semiconductor quantum dots \cite{thapa2025single}, colour centres in wide-bandgap materials such as diamond and silicon carbide \cite{herrmann2026highly, banasiak2026diamond, redhu2026bright, zhou2025silicon}, and defect emitters in two-dimensional materials such as hexagonal boron nitride and transition metal dichalcogenide monolayers \cite{boretti2025advancing, holewa2025solid, aharonovich2016solid, lyu2025tunable, raj2025single} . While these systems can in principle deliver single photons on demand, they typically require elaborate experimental infrastructure. Further, since the generation of the photon is related to an optical transition, these material platforms  have limited spectral range and less engineering flexibility, with most emitting outside the telecom bands. The probabilistic techniques generate photons at random through a quantum non-linear process. A photon pair is produced using either Spontaneous Parametric Down Conversion (SPDC) or Four Wave Mixing (FWM) and one of the photons is used to herald the production of the other photon \cite{fox2006quantum, amorim2026ultra,iveson2026fibre, kopylov2025spectral, lei2025efficient}. By using a nonlinear process together with precise waveguide engineering, the photons can be accurately generated at a desired wavelength.  

We here report on the realization of a heralded photon-pair source based on SPDC in an AlGaAs Bragg reflection waveguide that operates in the telecom C-band. AlGaAs is our preferred non-linear material because of its high $\chi^{(2)}$ value, room temperature operation, and a tunable transparency window \cite{baboux2023nonlinear}. The isotropic structure of the material makes it intrinsically non-birefringent, which eliminates the need for any walk-off compensation between the generated photons. Thus, by using a type II SPDC process where the signal and idler photons are orthogonally polarized with respect to each other, we have a source that produces inherently polarization-entangled photons without the need for any additional setup. The direct band gap of the material also enables the integration of a pump diode on the same chip. 

Since AlGaAs lacks birefringence, phase matching in these structures is achieved  by guiding the pump as a Bragg mode, whose effective index is set by the epitaxial layer stack such that it matches the index of the total-internal-reflection modes carrying the down-converted photons. Sources based on AlGaAs Bragg reflection waveguide (BRW) have advanced considerably over the past decade: polarization entanglement was generated directly from a monolithic BRW chip at telecom wavelengths without compensation, filtering or post-selection \cite{horn2013inherent}, and refinements of the epitaxial design and fabrication have since improved the spectral and temporal properties of the emitted pairs \cite{schlager2017temporally, chen2018high}. Electrically driven operation at room temperature \cite{boitier2014electrically} and pumping by a monolithically integrated quantum dot laser \cite{schlager2021difference} have likewise been demonstrated. More recently, AlGaAs sources have been taken to the network level, with multi-user quantum key distribution over fiber links of up to 50$\,$km, reaching up to eight users via BBM92 \cite{autebert2016multi, appas2021flexible}, a demonstration of quantum nonlocality in a three-party triangle network \cite{meskine2025experimental}, and a five-user quantum virtual local area network operated on a campus fiber network \cite{qian2025five}. These demonstrations have, however, largely relied on spooled fiber under laboratory conditions or on short campus links. Here we extend this to a 28$\,$km deployed urban fiber link, where the continuously varying birefringence of the installed fiber must be actively compensated, and assess how well the entanglement from a BRW source survives under such conditions.

 We characterize our source with regard to the quality of single photons by measuring photon pair rates, and the second order autocorrelation function ($g^{(2)}(\tau = 0)$). To evaluate the quality of entanglement we perform a quantum tomography measurement on the produced state and measure the purity and fidelity with respect to the $|\Psi^{+}\rangle$ Bell state. Finally, we distribute these entangled photons over a commercial fiber link of 28$\,$km length. The track is actively stabilized for polarization drifts and we reverify the entanglement between the distributed photons after transmission over the fiber.  

\section{\label{sec:level1}Device and Experimental setup}
\begin{figure*}[hbtp]
    \centering
    \includegraphics[width=\textwidth]{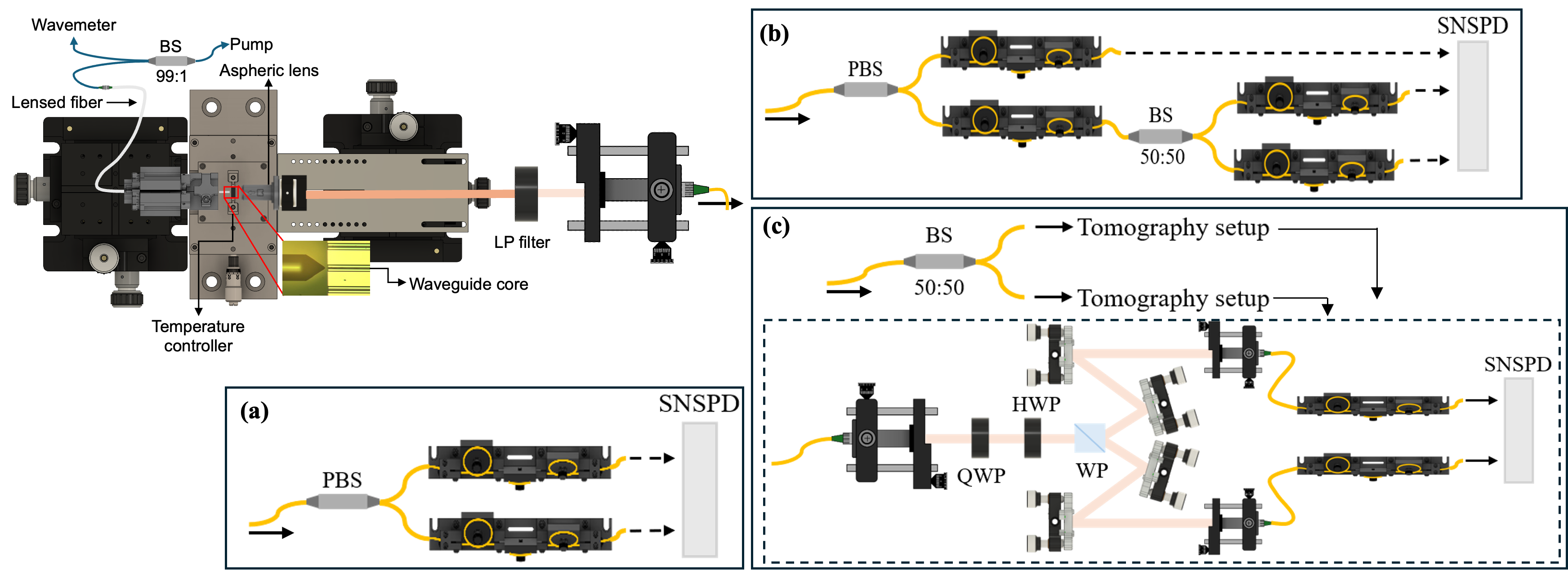}
    \caption{\label{fig:allsetup}The experimental setup used for different measurements.  The general setup to align the waveguide, is shown on the left. (a) Coincidence to accidental ratio measurement. (b) Second order correlation function measurement. (c) Entanglement measurement. A detailed explanation for each setup can be found in the text. Abbreviations: LP filter: long pass filter, PBS: polarizing beam splitter, BS: beam splitter, QWP: quarter–wave plate, HWP: half–wave plate, WP: wollaston prism, SNSPD: superconducting nanowire single–photon detector. }
\end{figure*}

 In our experiment we use AlGaAs ridge waveguides \cite{yang2020non} of  2$\,$mm length and 1.8$\,\mu$m ridge width to generate degenerate SPDC photons at 1544$\,$nm from pump photons at 772$\,$nm emitted by a cw laser (TOPTICA DL pro 780). The pump light is coupled into the waveguide using a polarization maintaining lensed fiber with a spot size of 2$\,\mu$m (OZ Optics). The waveguide is maintained at 20$^{\circ}$$\,$C for temperature phase matching ensuring efficient pair generation. The SPDC photons are then out-coupled using an uncoated aspheric lens (focal length = 4.51$\,$mm, NA = 0.55). A long pass filter at 1400$\,$nm is used to eliminate the residual pump light and the generated photons are coupled into a single mode fiber. All the measurement setups are shown in Fig. \ref{fig:allsetup}. For detection we use superconducting nanowire single photon detectors (SNSPD, ID Quantique, average detector efficiency:\,87\,\%, average dark count rate:\,16$\,$cps) and a time tagging device (ID900) for time-correlated single photon counting.  

\section{\label{sec:level1}Experimental results}
\subsection{\label{sec:level2}Photon pair generation}
\begin{figure}[hbtp]
    \centering
    \includegraphics[width=\columnwidth]{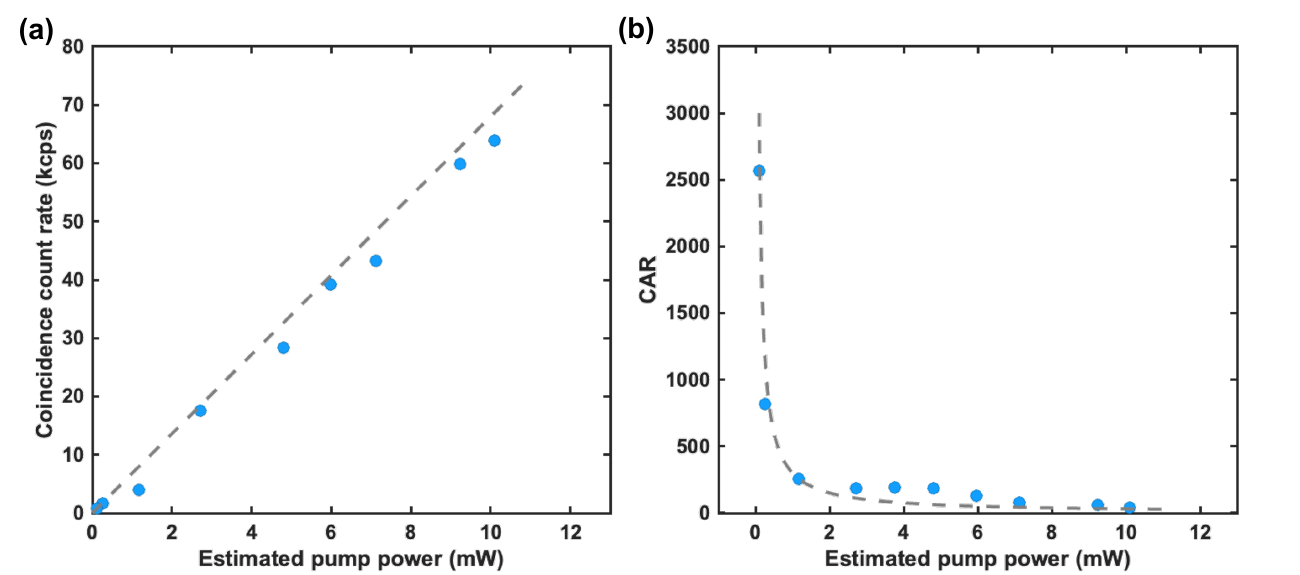}
    \caption{\label{fig:Coinc_CAR} (a) Measured raw coincidence counts as the function of estimated pump power at the tip of lensed fiber. (b) Coincidence-to-accidental ratio (CAR) variation as function of estimated pump power at the tip of lensed fiber. The dashed lines are to guide the eye. }
\end{figure}

We characterize the quality of our source by measuring an important figure of merit, the coincidence-to-accidental ratio (CAR). This is measured by deterministically separating the photon pairs with a fiber polarizing beam splitter (PBS), as shown in Fig. \ref{fig:allsetup}(a). The measurement shown in Fig. \ref{fig:Coinc_CAR} resembles a typical dependence of the coincidence rate and CAR on the pump power for SPDC-generated single photons in the low-pump regime. The quoted pump power corresponds to the estimated value at the tip of the lensed fiber. The estimation is based on the known transmission loss of the fiber and the measured power at the fiber input. As the pump power increases, the coincidence counts rise linearly while the CAR decreases inversely. This linear dependence is expected for SPDC in the low-gain regime, where the pair generation rate scales directly with the pump photon flux. However, the probability of producing multi-photon events increases rapidly than the single-pair rate leading to a drop in the overall CAR. At the maximum estimated pump power of 9.4$\,$mW, we measure $\approx$ 313$\,$kcps coincidence counts and a CAR of $\,$33.35. 

To compare our source to others, we measure its brightness in terms of the pair generation rate (PGR). PGR is defined as the rate of photons produced per unit of pump power. It is obtained from the net coincidence rate after correcting for the transmission of the detection setup. To measure this value, a variable attenuator is connected before the fiber PBS in the setup Fig. \ref{fig:allsetup}(a). The single counts in each detector and the coincidence counts are measured with varying attenuation levels. The data is then fitted with the method described in \cite{appas2022nonlinear} resulting in a PGR of $2.88\times10^{7}$$\,$/s/mW for our source. 

\subsection{\label{sec:level2}Single photon quality}
We measure the second order autocorrelation function at zero time delay ($g^{(2)}(0)$) to quantify the purity of the generated single photons. To this end we measure one photon of the pair as herald and use a Hanbury Brown and Twiss (HBT) setup to determine the other photon's purity \cite{fox2006quantum, Beck07}. 

\begin{figure}
    \centering
    \includegraphics[width=\columnwidth]{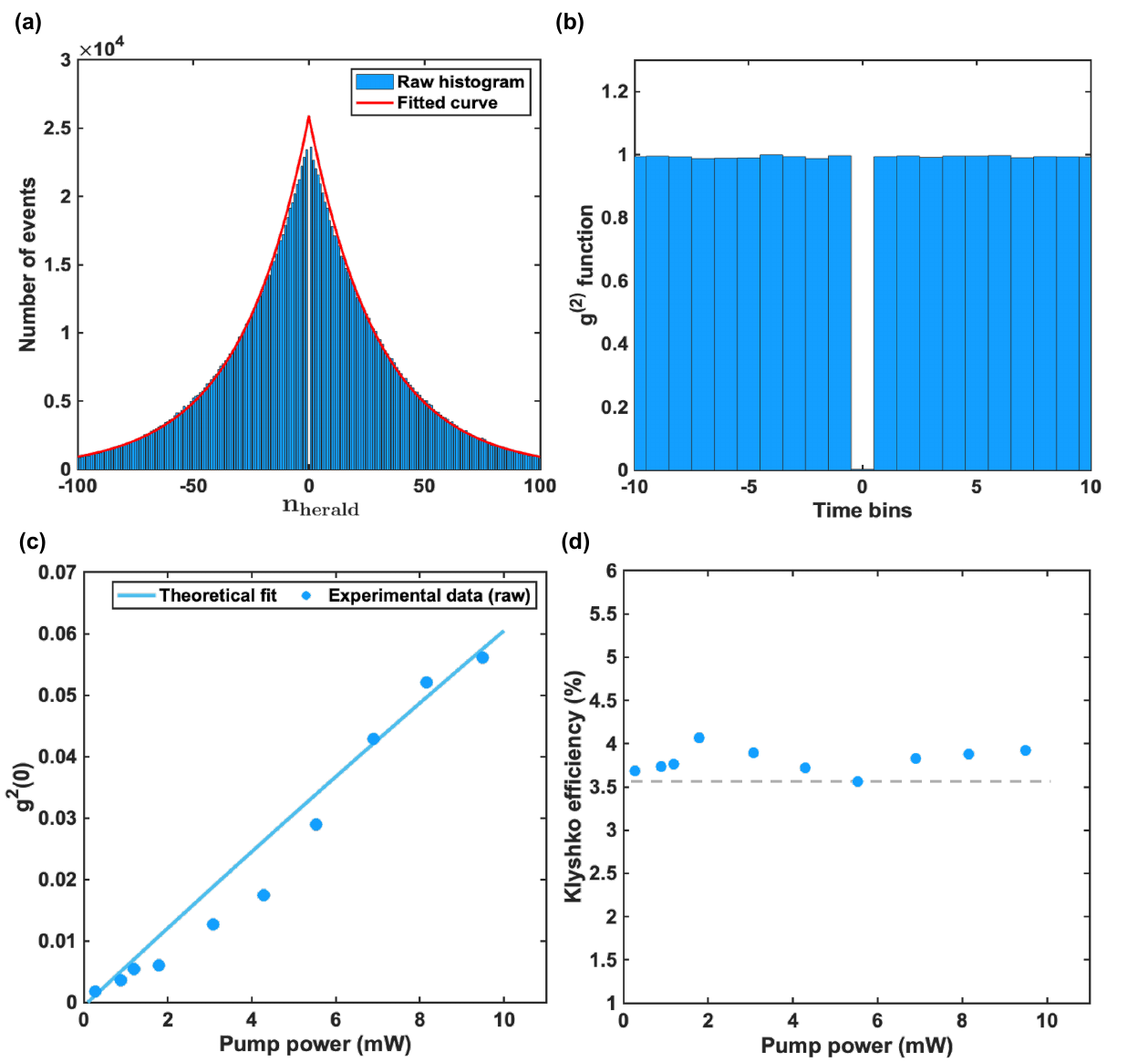}
    \caption{\label{fig:g2_combined} (a) Raw histogram of detection events between detector 2 and 3 extracted from the recorded timestamp. (b) Zoomed and normalized histogram around the central bin to infer $g^{(2)}(0)$. (c) Variation of $g^{(2)}(0)$ with estimated pump power at the tip of lensed fiber. The $g^{(2)}(0)$ values are extracted from the height of central bin in the histogram as in (b). (d) Variation of Klyshko efficiency with estimated pump power. The dashed line is to guide the eye. }
\end{figure}

The experimental setup is shown in Fig. \ref{fig:allsetup}(b). The generated photon pairs are first split deterministically using a fiber PBS. One of the photons from the pair (signal) is detected by a SNSPD (D1). The HBT measurement is performed on the idler photon from the same pair. The two SNSPD used in the HBT measurement, D2 and D3, are conditioned on heralding events registered by D1. The heralded second order correlation is then given by $g^{(2)}(0) = \frac{N_1 C_{123}}{C_{12} C_{13}}$ where $N_1$ is the single count rate on D1 and 
$C_{12},\,C_{13},\,C_{123}$ are the two and three fold coincidence counts between the detectors indicated by the subscripts \cite{Beck07}. In our case the photon pairs are created at random time because of continuous pumping and the spontaneous non-linear creation process. We use the method described in Ref. [\onlinecite{fasel2004high}] to compute the second order autocorrelation value from the raw binary data recorded by the time tagging device. The raw binary detection record is used to build a histogram $M(n)$ of the number of herald trigger cycles $n \equiv n_{\text{herald}}$ separating a click on one HBT detector (D2) from a click on the other HBT detector (D3), with $n=0$ corresponding to a simultaneous click on both HBT detectors within the same herald trigger. The wings of this histogram ($n \neq 0$) follow
\begin{equation}
    M(n) = C\, p_2 (1-p_3)^{|n|} p_3,
\end{equation}
where $p_2$ and $p_3$ are the per-trigger click probabilities of D2 and D3; fitting this decay yields the normalization constant $C$. Since the fitted decay accounts for the uncorrelated background, the normalized central bin yields $g^{(2)}(0)$ without requiring a separate subtraction of dark counts. The histogram of the number of heralding triggers ($n_{\text{herald}}$) between detection events at the HBT detectors is shown in Fig. \ref{fig:g2_combined}(a). The near zero central bin indicates that there was no simultaneous detection on both detectors for the same trigger. To estimate the value of second order correlation function, we restrict the region of interest around the central bin. Fig. \ref{fig:g2_combined}(b) shows the zoomed and normalized values from Fig. \ref{fig:g2_combined}(a) around the central bin. The height of the central bin in this figure directly indicates the value of $g^{(2)}(0)$. The photons show anti-bunching with $g^{(2)}(0) = 9.73 \pm 4.86 \times 10^{-5}$ at the estimated pump power of 0.0262$\,$mW at the tip of the lensed fiber. The measured $g^{(2)}(0)$ clearly indicates anti-bunching and marks successful generation of single photons. We repeat this measurement with varying pump powers. The results are shown in Fig. \ref{fig:g2_combined}(c). For each point in Fig. \ref{fig:g2_combined}(c) a histogram analogous to Fig. \ref{fig:g2_combined}(b) was measured and $g^{(2)}(0)$ was calculated from the height of central bin. The increasing values of $g^{(2)}(0)$ with pump power reflect the predicted nature of SPDC photons, as the multi-photon generation probability increases at higher power.  We also determine another figure of merit, i.e. the Klyshko efficiency defined as the ratio of idler pair rate to signal counts. These two figures of merit are largely independent, as $g^{(2)}(0)$ probes the source's photon-number statistics while the Klyshko efficiency quantifies linear channel loss, which does not affect the former. The variation of Klyshko efficiency with pump power is shown in Fig. \ref{fig:g2_combined}(d). The values indicate the probability of detecting a photon from the pair given that the other was detected as herald, and is thus independent of the pump power. We measure this end to end efficiency in our setup to be $\approx\,4\%$. Considerably higher heralding efficiencies have been reported for waveguide sources in other material platforms, for instance $64.1 \pm 2.1\,\%$ for a PPLN ridge waveguide source operating at telecom wavelengths \cite{bock2016highly}. The lower value obtained here is primarily attributed to mode-mismatch losses at the fiber-chip interface and to Fresnel reflection at the uncoated, high-index-contrast AlGaAs waveguide facets, rather than to the intrinsic pair-generation efficiency of the source. 

\subsection{\label{sec:level2}Entanglement measurement - local}

\begin{figure}
    \centering
    \includegraphics[width=\columnwidth]{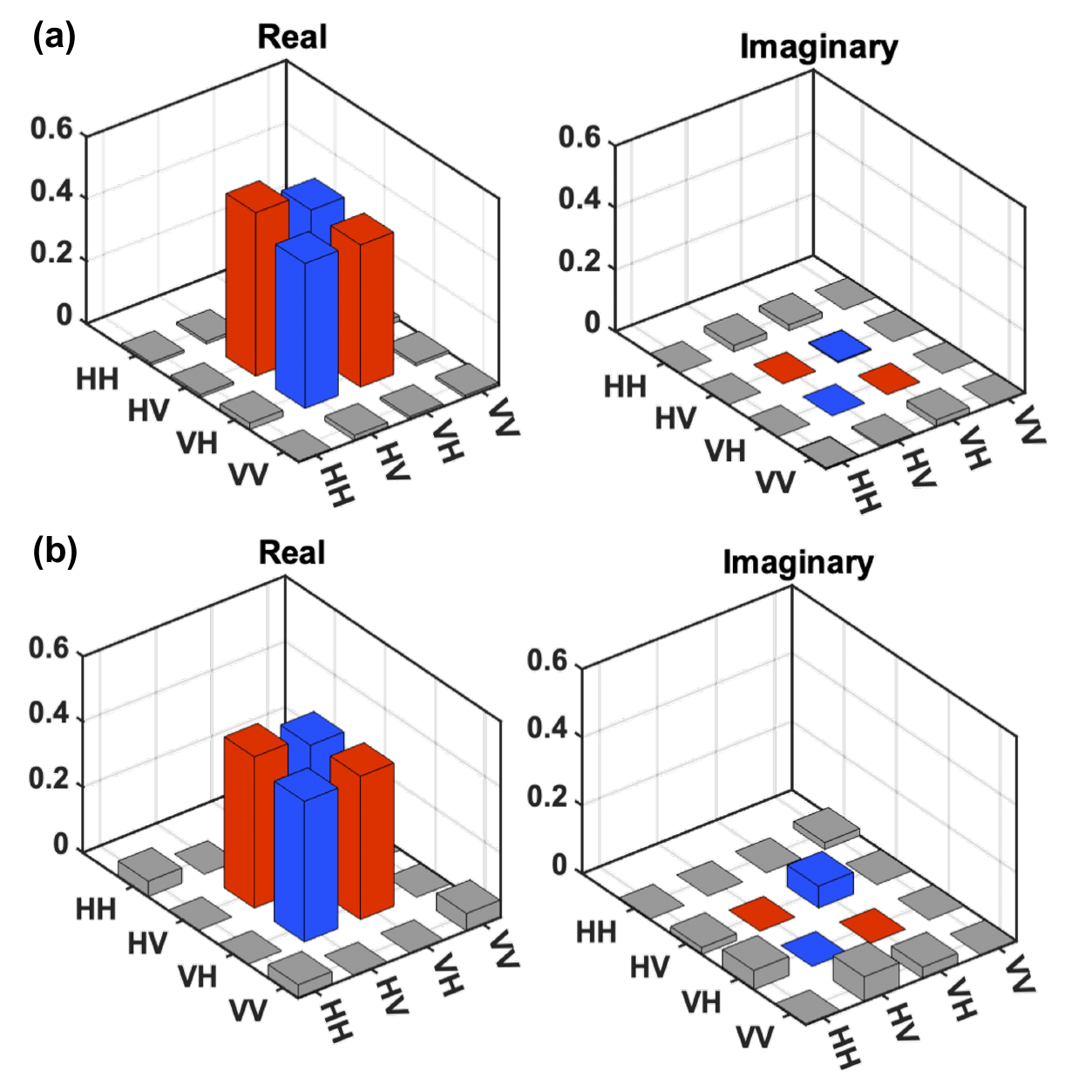}
    \caption{\label{fig:DensityMatrix_combined} Reconstructed density matrix using maximum likelihood estimation (a) local measurement. (b) fiber track measurement. The measurement is performed at an estimated pump power of 10$\,$mW incident on the waveguide facet.}
\end{figure}

Now that the generation of heralded single photons is established we proceed with verifying polarization entanglement. In a first step this entanglement measurement is performed locally by connecting the source to the measurement setups through a single mode fiber. The photons from the source are first filtered using a 10$\,$nm bandpass filter and then split non-deterministically using a 50:50 non-polarizing fiber beam splitter. Each output arm of the beam splitter is connected to the measurement setups as shown in Fig. \ref{fig:allsetup}(c). The photons pass sequentially through a quarter-wave plate, a half-wave plate, and then a Wollaston prism. The wave plates are rotated to a set of precise angles to realize different polarization basis settings. Dividing the input beam into two orthogonally polarized partial beams via the prisms facilitates simultaneous measurements on both polarizations. For each polarization basis, the count rate is measured, and the resulting data are then analyzed to extract the coincidence counts corresponding to that basis. The average CAR, calculated from the coincidence histogram of the basis with maximum expected coincidences (HV and VH basis projection) was 96.5. The coincidence counts from all measurement bases are subsequently used to reconstruct the density matrix of the measured state via maximum likelihood estimation \cite{james2001measurement}. The real and imaginary part of the reconstructed density matrix is shown in Fig. \ref{fig:DensityMatrix_combined}(a). We calculate a fidelity of 95.24(2)$\,\%$ and a purity of 91.38(3)$\,\%$ with respect to the $|\Psi^{+}\rangle$ Bell state. 

Since the polarization is rotated due to birefringence upon transmitting through a single mode fiber it is necessary to determine and compensate this rotation. To estimate the rotation matrix we send a set of polarization states with orthogonal Stokes vectors on the Poincaré sphere (horizontal: $|H\rangle$ and diagonal: $|D\rangle$) through the fiber to the measurement setups. This is done by placing a polarizer and half wave plate into our source setup. The measured Stokes vector is compared to the known Stokes vector of the input state to determine the rotation. The inverse of this measured rotation matrix is applied to the wave plate angles when measuring different polarization basis.  

\begin{figure}
    \centering
    \includegraphics[width=\columnwidth]{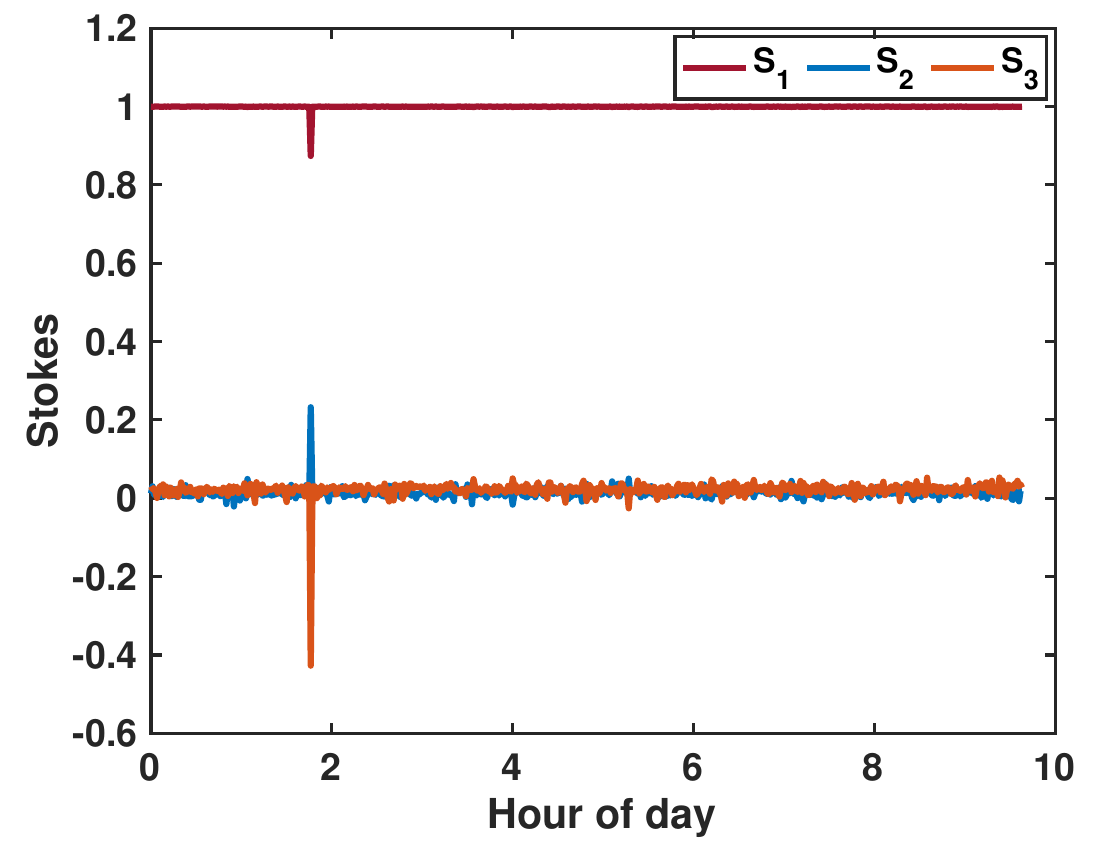}
    \caption{\label{fig:LocalStability} Polarization drift in the laboratory fiber. The normalized Stokes parameters $S_1$, $S_2$ and $S_3$ are recorded over time for an $\ket{H}$ polarized input, without active stabilization.}
\end{figure}

It is also important to ensure that this rotation is constant throughout the measurement time. We measure the stability of polarization over a duration of almost 10$\,$hours by sending horizontally polarized photons over the fiber. The measured Stokes vector in Fig. \ref{fig:LocalStability} confirmed the polarization stability.  Since horizontal polarization was used to test this stabilization, the $S_{1}$ Stokes vector appears more stable than $S_{2}$ and $S_{3}$ due to their greater sensitivity to shot noise.  The abrupt change just after 100$\,$minutes was caused by accidentally moving the fiber during the measurement.  Given the actual measurement time is less than 10$\,$hours, measuring the rotation matrix once before starting is sufficient.  \\

\subsection{\label{sec:level3}Entanglement distribution over fiber link}

\begin{figure}[hbtp]
    \centering
    \includegraphics[width=\columnwidth]{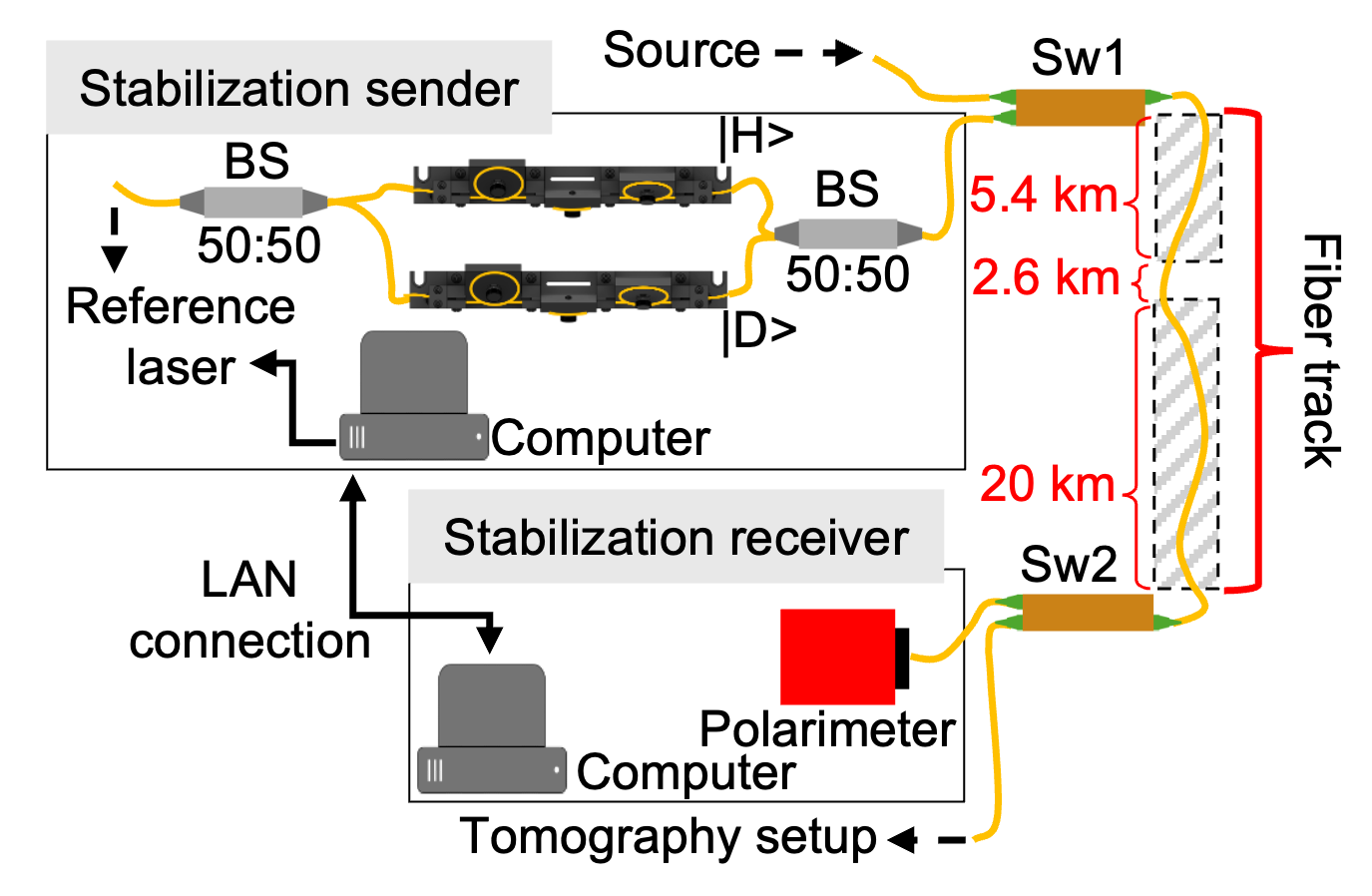}
    \caption{\label{fig:TrackStabilitySetup} Measurement setup for the fiber link. The shaded region of fiber track runs underground while the rest is deployed overground. Sw1 and Sw2 are fiber switches. }
\end{figure}

Next, we distribute the entangled photons over a commercial deployed telecom fiber link with a length of $28\,$km \cite{kucera2024demonstration}. This fiber link consists of two 14$\,$km long dark fibers running in parallel across the metropolitan area of Saarbr{\"u}cken to a second location where they are connected to a loop. The major part of the fiber link runs underground while 2.6$\,$Km in total are deployed along an aerial high power line. Due to lossy patch points, the total transmission is reduced to 1.26$\,\%$ (19$\,$dB loss). The average CAR, calculated from the coincidence histogram of HV and VH basis projection decreased to 11.76 due to low transmission and high background noise. The photons from the source are sent over to another laboratory through a single mode fiber going across building corridors. For simplicity, this laboratory will hereafter be referred to as lab II, while the laboratory in which the source is located will be designated as lab I throughout the remainder of the text. In lab II the photons are connected to the fiber track and sent back to lab I after transmission for detection and measurement in different polarization basis using the same setup shown in Fig. \ref{fig:allsetup}(c). 

\begin{figure}\label{fig:PolDrift}
    \centering
    \includegraphics[width=\columnwidth]{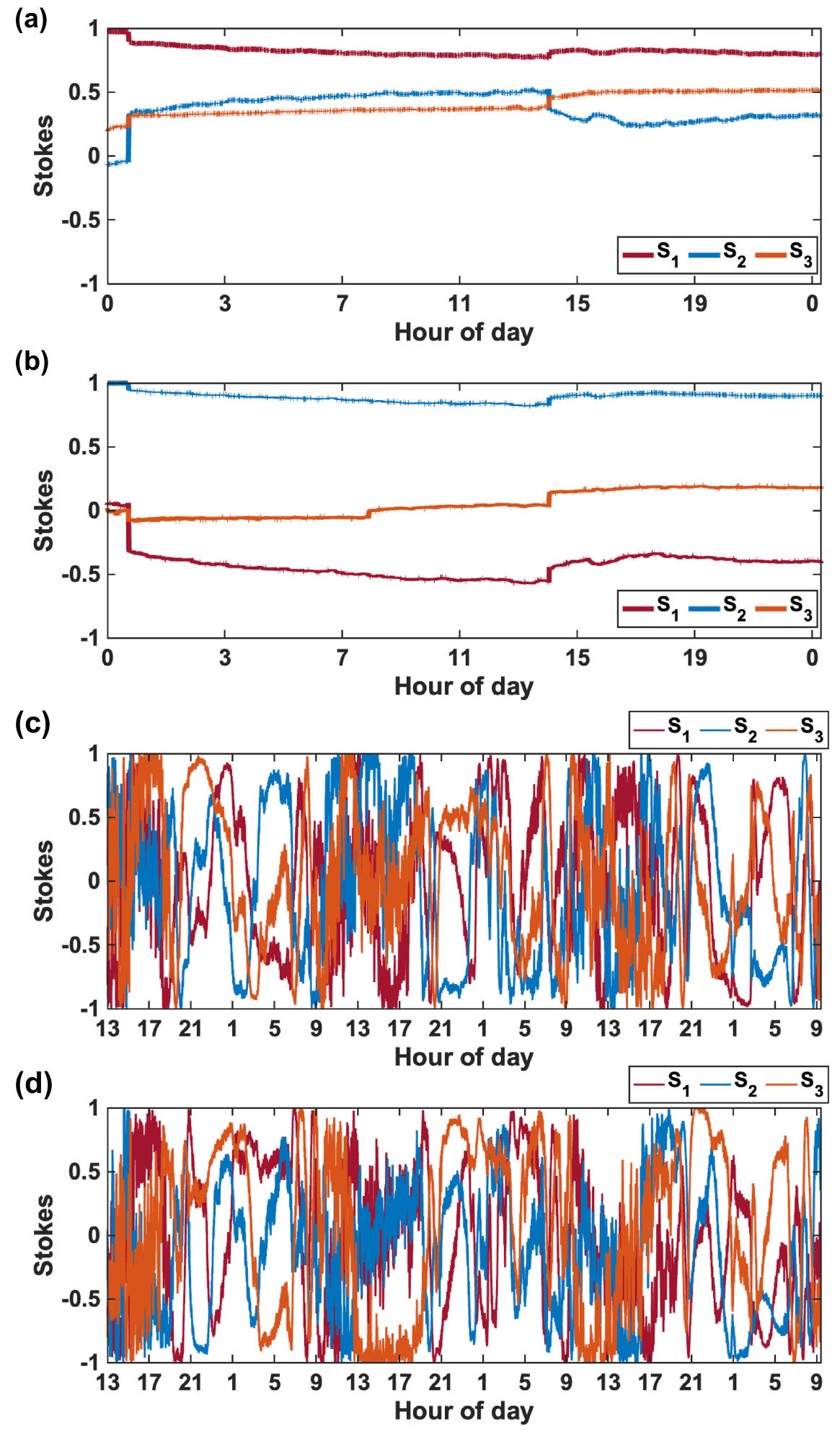}
   \caption{\label{fig:TrackPolStability_combined} Drift of polarization Stokes components over time. (a-b) Fiber between lab I and II for input polarizations $\ket{H}$ and $\ket{D}$ respectively, (c-d) $28\,$km long fiber track for input polarizations $\ket{H}$ and $\ket{D}$ respectively.}
\end{figure}

\begin{figure}
    \centering
    \includegraphics[width=\columnwidth]{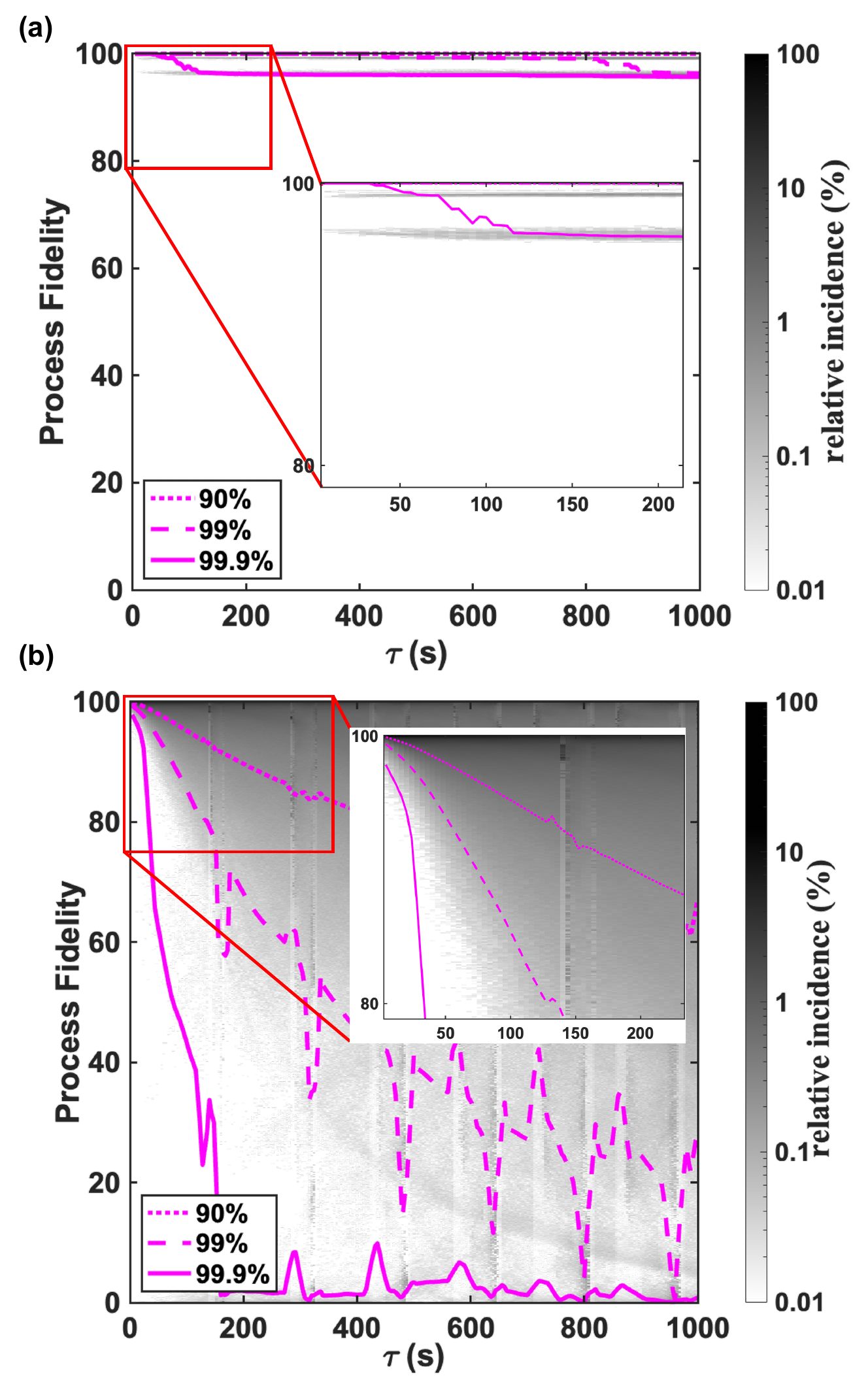}
   \caption{\label{fig:TrackPF_combined} Relative incidence (grayscale) to measure a process fidelity after a drift time $\tau$, derived from the data shown in Fig. \ref{fig:TrackPolStability_combined}. Contours of certain probability thresholds are displayed as purple lines. (a) Process fidelity on the fiber between lab I and II. (b) Process fidelity on the fiber track.}
\end{figure}

Similar to the local measurement, it is important to compensate polarization drifts. The polarization stabilization consists of two stages, first the drift compensation on the fiber track located in lab II and second the stabilization of the fibers between the two labs. To estimate the effect in both of these fibers, the Stokes parameters of transmitted polarized light with input polarizations $|H\rangle$ and $|D\rangle$ are measured over time, as shown in Fig. \ref{fig:TrackPolStability_combined}. We quantify the quality of polarization encoded quantum information via the process fidelity $\mathcal{F}_\mathrm{P}=\frac{1+\mathrm{Tr}(M)}{4}$, with the polarization rotation matrix $M$ \cite{kucera2024demonstration}. With this measurement, we extract from the polarization drift the drift time $\tau$ up to which we can expect a certain process fidelity with high probability. As seen in Fig. \ref{fig:TrackPolStability_combined}(a-b), the polarization on the fiber between labs drifts very slowly. For this section of the fiber a process fidelity of >95$\,$\% and >99$\,$\% is maintained with the probability of $99.9\,\%$ and $90\,\%$ respectively (Fig. \ref{fig:TrackPF_combined}(a)) up to the maximum measurement time 1000$\,$s. On the other hand, the polarization on fiber track is fluctuating rapidly and needs to be stabilized more often (Fig. \ref{fig:TrackPolStability_combined}(c-d)). To maintain 99\% process fidelity with a 90\% probability we stabilize the track every 5 seconds (Fig. \ref{fig:TrackPF_combined}(b)), using a piezo controller based setup described in Ref.[\onlinecite{kucera2024demonstration}]. The remaining polarization drifts on the local fibers are compensated every 30$\,$minutes using the wave plate based correction scheme described in \ref{sec:level2}.

\begin{figure}
    \centering
    \includegraphics[width=\columnwidth]{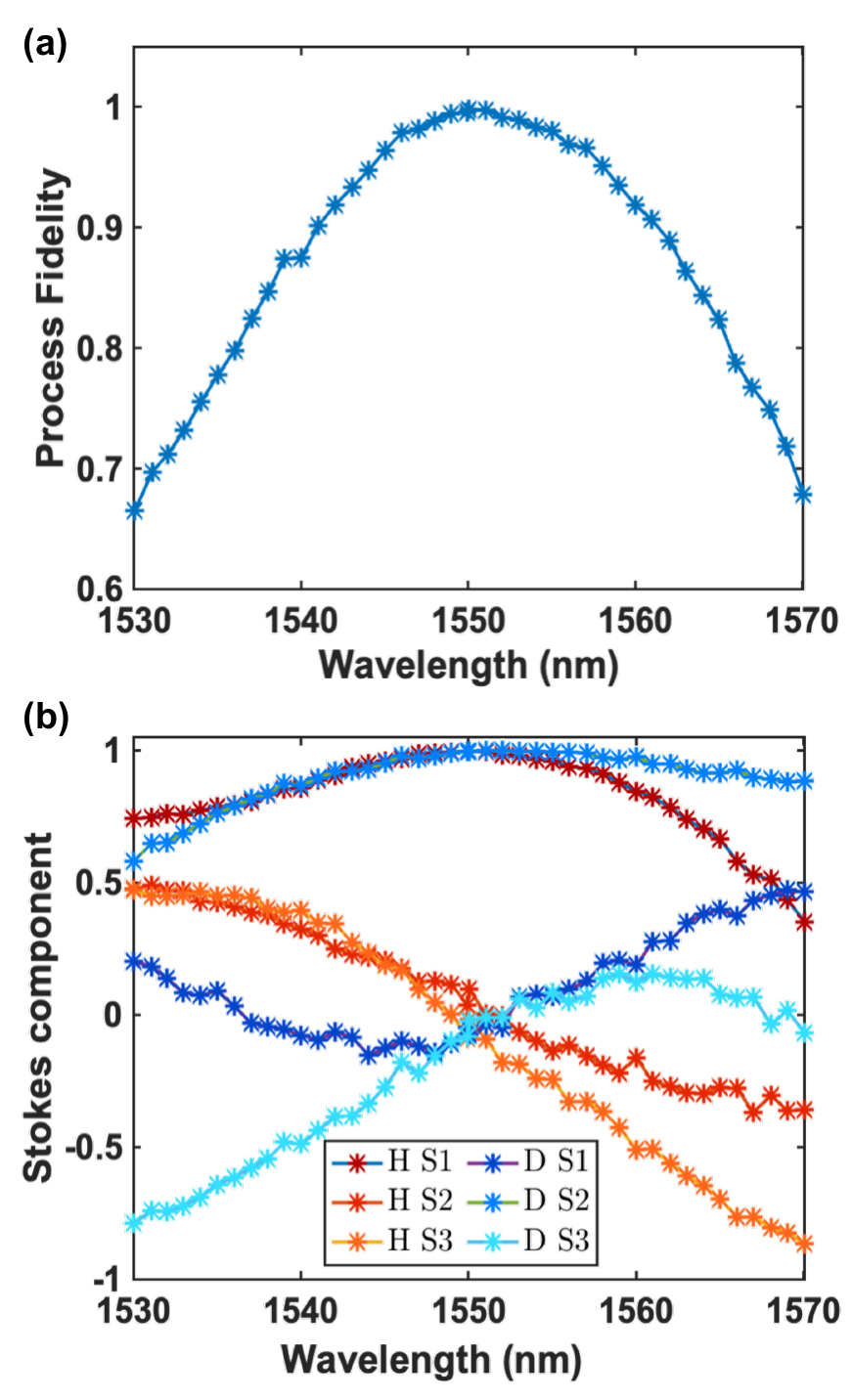}
    \caption{\label{fig:PFwithWavelength} Polarization transmission behavior dependent on wavelength on the one-way fiber track. The transmitted polarization is first set to the input polarizations $\ket{H}$ and $\ket{D}$ at a wavelength of $1550\,$nm using the polarization stabilization. Afterwards different wavelengths with the same input polarizations are transmitted. (a) Process fidelity compared to $1550\,$nm transmission. (b) Stokes components of the transmitted signal for both input polarizations. Plot adapted from \cite{Dick2022}.}
\end{figure}

Although the fiber link is actively stabilized, there can be some uncompensated polarization rotation due to difference in the wavelengths of the polarized reference light and the entangled photons. For the entanglement measurement, we filter light from the source using a 10$\,$nm bandpass filter, whereas the laser used to send a fixed polarization has a narrow line-width ($\approx$ 10-15$\,$kHz) close to $1550\,$nm. The effect of this wavelength mismatch on the process  fidelity was measured by first applying polarization compensation using a $1550\,$nm laser and subsequently transmitting light from a tunable laser. The resulting wavelength dependence of the process fidelity and the Stokes vector is shown in Fig. \ref{fig:PFwithWavelength} for a transmission over one of the $14\,$km fibers. With a 10$\,$nm bandpass filter creating an approximately uniform spectral distribution of the photon bandwidths, the estimated process fidelity compared to matching reference wavelength reduces to about $\mathcal{F}_\text{p}\approx 98.7\,\%$ for transmission over the full fiber track \cite{Dick2022}. 

Using tomography as before, we measure a fidelity of 90.26(84)$\,\%$ with the $|\Psi^{+}\rangle$ Bell state and a purity of 88.99(144)$\,\%$ after transmission over the actively stabilized fiber link. The real and imaginary part of the estimated density matrix is shown in Fig. \ref{fig:DensityMatrix_combined}(b). It can be seen that in comparison to the local measurement there is some non-zero imaginary components in the track measurement. This is because some additional polarization-dependent components (e.g. beam sampler, polarization paddle) are introduced during the measurement. These components may lead to some extra phase resulting in a non-zero imaginary part. The reduced fidelity is a result of contributions from several factors such as the lower signal-to-noise ratio on the track compared to the local measurement as well as the broad photon spectrum. Treating the background as white noise \cite{Werner_1989} and accounting for wavelength-dependent polarization rotations across the photon spectrum, the local fidelity, accounting for the local CAR, and the CAR of the loop measurement yield an estimated fidelity of $88.7(15)\%$, in agreement with the measured value within the uncertainties. Additionally, even with active compensation there can be some residual uncompensated polarization rotation that affects the fidelity of state.

\section{\label{sec:level1}Conclusion}
In conclusion, we present a room-temperature source of single photons and entangled photon pairs at telecom wavelengths, based on an AlGaAs Bragg reflection waveguides. The source reaches a pair generation rate of $2.88\times10^{7}$$\,$/s/mW. Such a high rate is important for distributing the entangled photons over deployed fiber, since the coincidence rate falls with the product of the transmissions of both links. Our source shows excellent anti-bunching with $g^{(2)}(\tau) = 9.73\times 10^{-5}$ at a pump power of 0.0262$\,$mW. The source generates a state with 95.24(2)$\,\%$ fidelity and 91.38(3)$\,\%$ purity with respect to the $|\psi^{+}\rangle$ Bell state. After distribution over an actively stabilized fiber link, the fidelity remains at 90.26(84)$\,\%$ and the purity at 88.99(144)$\,\%$. The small decrease is fully explained by the lower signal-to-noise ratio and residual polarization effects on the link. The entanglement therefore survives transmission through a deployed fiber link with only minor degradation, which makes our source well suited for practical quantum communication in fiber networks.\\
Looking ahead, this work points to several directions for further development. The end-to-end efficiency of the source is presently limited by the collection path rather than by the pair-generation process itself. Anti-reflection coating of the waveguide facets would remove the Fresnel losses associated with the high index contrast of AlGaAs, and adaptive mode matching between the guided mode and the spatial mode of the single-mode fiber would reduce the residual coupling losses. Since polarization drift is intrinsic to installed fiber, the active stabilization demonstrated here will be a necessary component of any practical deployment, and extending it to longer links and to several simultaneous users is a natural next step. The broad phase-matching bandwidth of the type-II process in AlGaAs Bragg reflection waveguides allows the emission spectrum to be divided into many energy-matched channel pairs, so that a single chip can serve several users simultaneously, while the planar geometry of the platform permits multiple waveguides to be patterned on the same chip, opening a route to spatial multiplexing. Combined with the prospect of monolithically integrating the pump laser on the same chip, this makes the AlGaAs Bragg reflection waveguide a compelling building block for compact, room-temperature sources in practical metropolitan quantum communication networks.

\begin{table*}
\caption{\label{tab:table1}Comparison with waveguide-based photon-pair sources. PGR is normalized to the pump power in the waveguide. CAR and PGR are as reported in each work and were obtained under different pump powers and filtering conditions.}
\begin{ruledtabular}
\begin{tabular*}{\textwidth}{@{\extracolsep{\fill}}lccccc}
Source & Entanglement & CAR & PGR (/s/mW) & Fidelity & Reference \\ \hline
AlGaAs-BRW\footnote{Bragg reflection waveguide.}
    & polarization & -- & $7.3\times10^{5}$ & 0.89 & \cite{kang2015two} \\
AlGaAs-BRW
    & polarization & -- & $1.2\times10^{7}$\footnote{Corrected for 80\% facet reflectivity; $2.4\times10^{6}$/s/mW without correction.}
    & $>0.96$\footnote{Over a 42\,nm bandwidth.} & \cite{JING2024100083} \\
AlGaAs-BRW
    & polarization & 80 & $3.4\times10^{6}$ & $>0.95$ & \cite{appas2022nonlinear} \\
AlGaAs counter-propagating
    & polarization--frequency & 42 & $3.3\times10^{5}$\footnote{$10^{7}$ pairs/s at 30\,mW pump.} & 0.85 & \cite{francesconi2023chip} \\
AlGaAs-on-insulator
    & -- & -- & $2.6\times10^{10}$ & -- & \cite{placke2024telecom} \\
TFLN (PPLN)\footnote{Thin-film lithium niobate.}
    & polarization & -- & $5.1\times10^{8}$ & 0.944 & \cite{kim2026integrated} \\
TFLN (PPLN)
    & energy--time & $>10^{5}$ & $1.3\times10^{10}$ & --\footnote{Two-photon interference visibility $>98\%$.} & \cite{PhysRevLett.127.183601} \\
This work (AlGaAs-BRW)
    & polarization & 33 & $2.88\times10^{7}$ & 0.952 & \\
\end{tabular*}
\end{ruledtabular}
\end{table*}

\begin{acknowledgments}
We gratefully acknowledge financial support from the Federal Ministry of Research, Technology and Space (BMFTR) through projects VOMBAT (16KIS1767), QR.N (16KIS2180) and TD.QR (16KISSO21K) and from the "Transformationsprogramm Forschung und Wissenstransfer Saar” through the Center for Quantum Technologies (QuTe). We are grateful to Damian Weber from htw saar for providing necessary lab infrastructure, to Dieter Klein and colleagues of HIZ Saarland for indispensable technical support with the fiber link and communication infrastructure, and to VSE NET GmbH for generously providing the fiber link.
\end{acknowledgments}

\section*{Data Availability Statement}
Data available on reasonable  request from the authors.

The data that support the findings of this study are available from the corresponding author upon reasonable request.

\bibliography{aipsamp}

@PREAMBLE{
  "\providecommand{\noopsort}[1]{}"
  # "\providecommand{\singleletter}[1]{#1}"
}

@ARTICLE{couteau2023applications,
  author="Christophe Couteau and Stefanie Barz and Thomas Durt and Thomas Gerrits and Jan Huwer and Robert Prevedel and John Rarity and Andrew Shields and Gregor Weihs",
  title="Applications of single photons to quantum communication and computing",
  journal="Nature Reviews Physics",
  volume="5",
  number="6",
  pages="326--338",
  year="2023",
}

@ARTICLE{thapa2025single,
  author="Dev Kumar Thapa and Soumava Biswas",
  title="Single photon generation from quantum dots: recent advances, challenges and future directions",
  journal="Nanoscale",
  volume="17",
  number="13",
  pages="7651--7672",
  year="2025",
}

@ARTICLE{tissot2026hybrid,
  author="Benedikt Tissot and Soubhadra Maiti and Emil R. Hellebek and Anders S{\o}ndberg S{\o}rensen",
  title="Hybrid single-ion atomic-ensemble node for high-rate remote entanglement generation",
  journal="Physical Review A",
  volume="114",
  number="1",
  pages="L010602",
  year="2026",
}

@ARTICLE{liu2026long,
  author="Wen-Zhao Liu and Ya-Bin Zhou and Jiu-Peng Chen and Bin Wang and Ao Teng and Xiao-Wen Han and Guang-Cheng Liu and Zhi-Jiong Zhang and Yi Yang and Feng-Guang Liu and others",
  title="Long-lived remote ion--ion entanglement for scalable quantum repeaters",
  journal="Nature",
  volume="652",
  number="8108",
  pages="51--57",
  year="2026",
}

@ARTICLE{wang2025scalable,
  author="Hui Wang and Timothy C. Ralph and Jelmer J. Renema and Chao-Yang Lu and Jian-Wei Pan",
  title="Scalable photonic quantum technologies",
  journal="Nature Materials",
  volume="24",
  number="12",
  pages="1883--1897",
  year="2025",
}

@ARTICLE{zhou2025silicon,
  author="Yu Zhou and Junhua Tan and HaiBo Hu and Sikai Hua and Chunhui Jiang and Bo Liang and Tongyuan Bao and Xinfang Nie and Shumin Xiao and Dawei Lu and Junfeng Wang and Qinghai Song",
  title="Silicon carbide: A promising platform for scalable quantum networks",
  journal="Applied Physics Reviews",
  volume="12",
  number="3",
  pages="031301",
  year="2025",
}

@ARTICLE{horn2013inherent,
  author="R. T. Horn and P. Kolenderski and D. Kang and P. Abolghasem and C. Scarcella and A. {Della Frera} and A. Tosi and L. G. Helt and S. V. Zhukovsky and J. E. Sipe and G. Weihs and A. S. Helmy and T. Jennewein",
  title="Inherent polarization entanglement generated from a monolithic semiconductor chip",
  journal="Scientific Reports",
  volume="3",
  pages="2314",
  year="2013",
}

@ARTICLE{schlager2017temporally,
  author="A. Schlager and B. Pressl and K. Laiho and H. Suchomel and M. Kamp and S. H{\"o}fling and C. Schneider and G. Weihs",
  title="Temporally versatile polarization entanglement from {B}ragg-reflection waveguides",
  journal="Optics Letters",
  volume="42",
  number="11",
  pages="2102--2105",
  year="2017",
}

@ARTICLE{chen2018high,
  author="Huan Chen and Silke Auchter and Maximilian Prilm{\"u}ller and Alexander Schlager and Thomas Kauten and Kaisa Laiho and Benedikt Pressl and Holger Suchomel and Martin Kamp and Sven H{\"o}fling and Christian Schneider and Gregor Weihs",
  title="Time-bin entangled photon pairs from {B}ragg-reflection waveguides",
  journal="APL Photonics",
  volume="3",
  number="8",
  pages="080804",
  year="2018",
}

@ARTICLE{boitier2014electrically,
  author="Fabien Boitier and Adeline Orieux and Claire Autebert and Aristide Lema{\^i}tre and Elisabeth Galopin and Christophe Manquest and Carlo Sirtori and Ivan Favero and Giuseppe Leo and Sara Ducci",
  title="Electrically injected photon-pair source at room temperature",
  journal="Physical Review Letters",
  volume="112",
  number="18",
  pages="183901",
  year="2014",
}

@ARTICLE{appas2021flexible,
  author="F{\'e}licien Appas and Florent Baboux and Maria I. Amanti and Aristide Lema{\^i}tre and Fabien Boitier and Eleni Diamanti and Sara Ducci",
  title="Flexible entanglement-distribution network with an {AlGaAs} chip for secure communications",
  journal="npj Quantum Information",
  volume="7",
  pages="118",
  year="2021",
}

@ARTICLE{meskine2025experimental,
  author="O. Meskine and I. {\v S}upi{\'{c}} and D. Markham and F. Appas and F. Boitier and M. Morassi and A. Lema{\^i}tre and M. I. Amanti and F. Baboux and E. Diamanti and S. Ducci",
  title="Experimental fiber-based quantum triangle-network nonlocality with a telecom {AlGaAs} multiplexed entangled-photon source",
  journal="PRX Quantum",
  volume="6",
  number="2",
  pages="020313",
  year="2025",
}

@ARTICLE{autebert2016multi,
  author="C. Autebert and J. Trapateau and A. Orieux and A. Lema{\^i}tre and C. Gomez-Carbonell and E. Diamanti and I. Zaquine and S. Ducci",
  title="Multi-user quantum key distribution with entangled photons from an {AlGaAs} chip",
  journal="Quantum Science and Technology",
  volume="1",
  number="1",
  pages="{01LT02}",
  year="2016",
}

@ARTICLE{qian2025five,
  author="Cheng Qian and Hongkai Tian and Xu Jing and Yichen Liu and Zhe Chen and Huiran Luo and Yuan Du and Xiaodong Zheng and Tangsheng Chen and Yuechan Kong and Hua-Lei Yin and Dong Jiang and Bin Niu and Liangliang Lu",
  title="Five-user quantum virtual local area network with an {AlGaAs} entangled photon source",
  journal="Science China Physics, Mechanics \& Astronomy",
  volume="68",
  number="3",
  pages="230311",
  year="2025",
}

@ARTICLE{schlager2021difference,
  author="A. Schlager and M. G{\"o}tsch and R. J. Chapman and S. Frick and H. Thiel and H. Suchomel and M. Kamp and S. H{\"o}fling and C. Schneider and G. Weihs",
  title="Difference-frequency generation in an {AlGaAs} {B}ragg-reflection waveguide using an on-chip electrically-pumped quantum dot laser",
  journal="Journal of Optics",
  volume="23",
  number="8",
  pages="085802",
  year="2021",
}

@ARTICLE{kopylov2025spectral,
  author="Denis A. Kopylov and Michael Stefszky and Torsten Meier and Christine Silberhorn and Polina R. Sharapova",
  title="Spectral and temporal properties of type-{II} parametric down-conversion: The impact of losses during state generation",
  journal="Physical Review Research",
  volume="7",
  number="3",
  pages="033122",
  year="2025",
}

@ARTICLE{yang2020non,
  author="Quankui Yang and Thorsten Passow",
  title="Non-ideal quarter-wavelength {B}ragg-reflection waveguides for nonlinear interaction: eigen equation and tolerance",
  journal="Optics Letters",
  volume="45",
  number="17",
  pages="4742--4745",
  year="2020",
}

@ARTICLE{lei2025efficient,
  author="Zhengshun Lei and Weiqiang Xie and Wenqi Wei and Zihao Wang and Ting Wang and Chao Xiang and Jianjun Zhang and Yikai Su",
  title="Efficient broadband wavelength conversion in {AlGaAsOI} nonlinear nanowaveguides using low-power continuous-wave pump",
  journal="ACS Photonics",
  volume="12",
  number="3",
  pages="1619--1627",
  year="2025",
}

@ARTICLE{amorim2026ultra,
  author="Rodrigo Amorim and Tim J. W{\"o}rmann and Andrius Zukauskas and Lars Gr{\"u}ner-Nielsen and Valdas Pasiskevicius and Karsten Rottwitt",
  title="Ultra-low-noise heralded single photons from counter-propagating {SPDC}",
  journal="Optica Quantum",
  volume="4",
  number="4",
  pages="382--388",
  year="2026",
}

@ARTICLE{iveson2026fibre,
  author="Peter Iveson and Paolo L. Mennea and Goronwy Tawy and Rex H. S. Bannerman and Noe Palomar-Davidson and Lewis D. Wright and Patrick M. Ledingham and Peter G. R. Smith and James C. Gates and Corin B. E. Gawith",
  title="Fibre-integrated {SPDC} heralded single-photon source using {Zn}-indiffused {PPLN} ridge waveguides at telecom wavelengths",
  journal="Optics Continuum",
  volume="5",
  number="9",
  pages="3079--3090",
  year="2026",
}

@ARTICLE{raj2025single,
  author="Akriti Raj and Swapnasagar Sahu and Peddarapu Lakshmi Narayani and Lakshminarayan Sharma and Siddharth Banerjee and Laxmi Narayan Tripathi",
  title="Single photon sources from {2D}-materials and interfacing with waveguides for quantum technological applications",
  journal="Advanced Quantum Technologies",
  volume="8",
  number="9",
  pages="2400648",
  year="2025",
}

@ARTICLE{lyu2025tunable,
  author="Xiaodan Lyu and Leevi Kallioniemi and Hao Hong and Rui Qu and Yan Zhang and Jesus Z{\'u}{\~n}iga-Perez and Kaihui Liu and Weibo Gao",
  title="A tunable entangled photon-pair source based on a van der {W}aals insulator",
  journal="Nature Communications",
  volume="16",
  number="1",
  pages="1899",
  year="2025",
}

@ARTICLE{banasiak2026diamond,
  author="Mariusz Banasiak and Mattia Pierpaoli and Robert Bogdanowicz",
  title="Diamond tin-vacancy centers: From synthesis and characterization to quantum photonic applications---a comprehensive review",
  journal="Materials Today Physics",
  volume="63",
  pages="102088",
  year="2026",
}

@ARTICLE{redhu2026bright,
  author="Ashish Redhu and Rajesh V. Nair",
  title="Bright single-photon emission from nanowaveguide-coupled {N-V} centers",
  journal="Physical Review Applied",
  volume="26",
  number="1",
  pages="014073",
  year="2026",
}

@ARTICLE{herrmann2026highly,
  author="Dennis Herrmann and Robert Morsch-Golsong and Tobias Bauer and Marlon Sch{\"a}fer and David Lindler and Linus Ehre and Peter van Loock and Matthew Markham and Nicola Palmer and Soumen Mandal and others",
  title="Highly indistinguishable photons from a tin-vacancy spin qubit in diamond",
  journal="arXiv preprint arXiv:2607.22439",
  year="2026",
}

@ARTICLE{holewa2025solid,
  author="Pawe{\l} Holewa and Andreas Reiserer and Tobias Heindel and Stefano Sanguinetti and Alexander Huck and Elizaveta Semenova",
  title="Solid-state single-photon sources operating in the telecom wavelength range",
  journal="Nanophotonics",
  volume="14",
  number="11",
  pages="1729--1774",
  year="2025",
}

@ARTICLE{boretti2025advancing,
  author="Alberto Boretti and Jonathan Blackledge and Stefania Castelletto",
  title="Advancing hexagonal boron nitride single photon sources: A strategic roadmap for quantum applications",
  journal="Materials Science in Semiconductor Processing",
  volume="185",
  pages="108932",
  year="2025",
}

@ARTICLE{aharonovich2016solid,
  author="Igor Aharonovich and Dirk Englund and Milos Toth",
  title="Solid-state single-photon emitters",
  journal="Nature Photonics",
  volume="10",
  number="10",
  pages="631--641",
  year="2016",
}

@BOOK{fox2006quantum,
  author="Anthony Mark Fox",
  title="{Quantum Optics: An Introduction}",
  volume="15",
  year="2006",
  publisher="Oxford University Press",
}

@ARTICLE{Beck07,
  author="M. Beck",
  title="Comparing measurements of {$g^{(2)}(0)$} performed with different coincidence detection techniques",
  journal="J. Opt. Soc. Am. B",
  volume="24",
  number="12",
  pages="2972--2978",
  month="Dec",
  year="2007",
}

@ARTICLE{baboux2023nonlinear,
  author="F. Baboux and G. Moody and S. Ducci",
  title="Nonlinear integrated quantum photonics with {AlGaAs}",
  journal="Optica",
  volume="10",
  number="7",
  pages="917--931",
  year="2023",
}

@ARTICLE{fasel2004high,
  author="Sylvain Fasel and Olivier Alibart and Sebastien Tanzilli and Pascal Baldi and Alexios Beveratos and Nicolas Gisin and Hugo Zbinden",
  title="High-quality asynchronous heralded single-photon source at telecom wavelength",
  journal="New Journal of Physics",
  volume="6",
  number="1",
  pages="163",
  year="2004",
}

@ARTICLE{james2001measurement,
  author="Daniel F. V. James and Paul G. Kwiat and William J. Munro and Andrew G. White",
  title="Measurement of qubits",
  journal="Physical Review A",
  volume="64",
  number="5",
  pages="052312",
  year="2001",
}

@ARTICLE{kucera2024demonstration,
  author="Stephan Kucera and Christian Haen and Elena Arensk{\"o}tter and Tobias Bauer and Jonas Meiers and Marlon Sch{\"a}fer and Ross Boland and Milad Yahyapour and Maurice Lessing and Ronald Holzwarth and others",
  title="Demonstration of quantum network protocols over a 14-km urban fiber link",
  journal="npj Quantum Information",
  volume="10",
  number="1",
  pages="88",
  year="2024",
}

@ARTICLE{appas2022nonlinear,
  author="F{\'e}licien Appas and Othmane Meskine and Aristide Lema{\^i}tre and Martina Morassi and Florent Baboux and Maria I. Amanti and Sara Ducci",
  title="Nonlinear quantum photonics with {AlGaAs} {B}ragg-reflection waveguides",
  journal="Journal of Lightwave Technology",
  volume="40",
  number="23",
  pages="7658--7667",
  year="2022",
}

@ARTICLE{francesconi2023chip,
  author="S. Francesconi and A. Raymond and R. Duhamel and P. Filloux and A. Lema{\^i}tre and P. Milman and Maria I. Amanti and F. Baboux and Sara Ducci",
  title="On-chip generation of hybrid polarization-frequency entangled biphoton states",
  journal="Photonics Research",
  volume="11",
  number="2",
  pages="270--278",
  year="2023",
}

@MASTERSTHESIS{Dick2022,
  author="Gabriel Gero Dick",
  title="Polarisationsstabilisierung einer Telekom-Faserstrecke",
  type="Bachelorarbeit",
  school="Universit{\"a}t des Saarlandes",
  address="Saarbr{\"u}cken",
  year="2022",
}

@ARTICLE{kang2015two,
  author="Dongpeng Kang and Minseok Kim and Haoyu He and Amr S. Helmy",
  title="Two polarization-entangled sources from the same semiconductor chip",
  journal="Physical Review A",
  volume="92",
  pages="013821",
  year="2015",
}

@ARTICLE{placke2024telecom,
  author="Marlon Placke and Jan Schlegel and Felix Mann and Pietro {Della Casa} and Andreas Thies and Markus Weyers and G{\"u}nther Tr{\"a}nkle and Sven Ramelow",
  title="Telecom-band spontaneous parametric down-conversion in {AlGaAs}-on-insulator waveguides",
  journal="Laser \& Photonics Reviews",
  volume="18",
  pages="2301293",
  year="2024",
}

@ARTICLE{kim2026integrated,
  author="Changhyun Kim and Hansol Kim and Minho Choi and Junhyung Lee and Yongchan Park and Sunghyun Moon and Jinil Lee and Hyeon Hwang and Min-Kyo Seo and Yoon-Ho Kim and Yong-Su Kim and Hojoong Jung and Hyounghan Kwon",
  title="Integrated bright source of polarization-entangled photons using lithium niobate photonic chips",
  journal="Optica",
  volume="13",
  pages="1553--1562",
  year="2026",
}

@ARTICLE{JING2024100083,
  author="Xu Jing and Cheng Qian and Xiaodong Zheng and Hu Nian and Chenquan Wang and Jie Tang and Xiaowen Gu and Yuechan Kong and Tangsheng Chen and Yichen Liu and Chong Sheng and Dong Jiang and Bin Niu and Liangliang Lu",
  title="Coexistence of multiuser entanglement distribution and classical light in optical fiber network with a semiconductor chip",
  journal="Chip",
  volume="3",
  number="2",
  pages="100083",
  year="2024",
}

@ARTICLE{PhysRevLett.127.183601,
  author="Usman A. Javid and Jingwei Ling and Jeremy Staffa and Mingxiao Li and Yang He and Qiang Lin",
  title="Ultrabroadband entangled photons on a nanophotonic chip",
  journal="Physical Review Letters",
  volume="127",
  number="18",
  pages="183601",
  year="2021",
}

@ARTICLE{Werner_1989,
  author="Reinhard F. Werner",
  title="Quantum states with {E}instein-{P}odolsky-{R}osen correlations admitting a hidden-variable model",
  journal="Physical Review A",
  volume="40",
  number="8",
  pages="4277--4281",
  year="1989",
}

@ARTICLE{bock2016highly,
  author="Matthias Bock and Andreas Lenhard and Christopher Chunnilall and Christoph Becher",
  title="Highly efficient heralded single-photon source for telecom wavelengths based on a {PPLN} waveguide",
  journal="Optics Express",
  volume="24",
  number="21",
  pages="23992--24001",
  year="2016",
}

\end{document}